\documentclass[useAMS,usenatbib]{mn2e}

\usepackage{epsfig}
\usepackage{longtable}
\usepackage{times} 
\usepackage{amsmath}
\def \apj {ApJ}
\def \apjl {ApJL}
\def \aap {A\&A}

\def \mnras {MNRAS}

\begin{document}

\title[Long term decay of the cyclotron line in Vela X-1]
{The Swift-BAT monitoring reveals a long term decay of the cyclotron line 
energy in Vela X-1}

\author[La Parola et al.]{V.\ La Parola$^{1}$, G.\ Cusumano$^{1}$, 
A.\ Segreto $^{1}$, A.\ D'A\`i$^{1}$\\
$^{1}$INAF, Istituto di Astrofisica Spaziale e Fisica Cosmica,
        Via U.\ La Malfa 153, I-90146 Palermo, Italy\\
}

\date{}

\pagerange{\pageref{firstpage}--\pageref{lastpage}} \pubyear{}

\maketitle

\label{firstpage}

\begin{abstract}

We study the behaviour of the cyclotron resonant scattering feature (CRSF) of 
the high mass 
X-ray binary Vela X-1 using the long-term hard X-ray monitoring performed by 
the Burst Alert Telescope (BAT) on board Swift. 
High statistics, intensity selected spectra were built along 11 years of
BAT survey. While the fundamental line is not revealed, the second harmonic of 
the CRSF can be clearly detected in
all the spectra, at an energy varying between $\sim 53$ keV and $\sim 58$ keV,
directly correlated with the luminosity. We have further investigated the
evolution of the CRSF in time, by studying the intensity selected spectra built
along four 33-month time intervals along the survey. For the first time we find
in this source a secular variation in the CRSF energy: independent
of the source luminosity, the CRSF 
second harmonic energy decreases by $\sim 0.36$ keV/year between the first and 
the third time interval, corresponding to an apparent decay of the magnetic 
field of $\sim 3\times 10^{10}$ G/year.  The intensity-cyclotron energy pattern 
is consistent between the third and the last time intervals. 
A possible interpretation for this decay could be the settling of an accreted 
mound that produces either a distortion of the poloidal magnetic field on the 
polar cap or a geometrical displacement of the line forming region. This 
hypothesis seems supported by the correspondance between the rate of the line shift 
per unit accreted mass and the mass accreted on the polar cap per unit area in
Vela X-1 and Her X-1, respectively.                                                 

\end{abstract}

\begin{keywords}
X-rays: binaries -- X-rays: individual: Vela X-1. 

\noindent
Facility: {\it Swift}

\end{keywords}


        \section{Introduction\label{intro}}

Vela X-1 is a wind-accreting neutron star with a spin period 
of $\sim283$\,s \citep{mcclintock76}, rotating
in an 8.9\,d \citep{vankerkwijk95} orbit around the B0.5Ib super-giant HD\,77523 
\citep{hiltner72}, at a distance $1.9\pm0.2$\,kpc \citep{sadakane85}. 
The mean luminosity of the source is $5\times10^{36}$\,erg\,s$^{-1}$ 
\citep{furst10}, and the flux shows a great variability even at short time
scales (hours), with the source going 
from off-states to giant flares up to a few Crab \citep{kreykenbohm08}.

Neutron stars in high mass X-ray binaries are characterized by intense magnetic
fields ($\sim 10^{12}$ G)  that in several sources can be directly measured 
through the energy of characteristic absorption lines called
cyclotron resonant scattering features (CRSF). A strong magnetic field causes  
the motion  of the electrons perpendicular to the magnetic field in 
the accreting plasma to be quantized in 
discrete Landau levels and photons with energies corresponding to these levels 
undergo resonant scattering producing an absorption feature in their spectra.

In Vela X-1 an absorption line feature at 55 keV was first reported by
\citet{kendziorra92} with data from the High Energy X-ray Experiment (HEXE), while
\citet{makishima92,choi96} reported an absorption feature at lower energy ($\sim
30$ keV) from Ginga data. Later
observations failed to detect this low energy feature, revealing only one
absorption line at $53\sim55$ keV \citep{orlandini98,labarbera03,kreykenbohm08}.
\citet{kretschmar97}, with HEXE, \citet{kreykenbohm02} with RXTE, 
\citet{schanne07} with INTEGRAL, and \citet{maitra13} and \citet{odaka13} with 
Suzaku confirmed the presence of both features, with energies varying between
$\sim 23$ keV and $\sim 27$ keV and between $\sim 45$ keV and $\sim 54$ keV,
respectively. The low energy feature, interpreted as the fundamental CRSF, 
is  considerably weaker than its second harmonic. \citet{furst14} analysed 
two NuSTAR observations with different luminosity levels detecting both 
features, but while the harmonic is always highly significant, the fundamental 
is barely detectable in the observation at higher intensity. They also show that
the depth of the two lines are anti-correlated and they interpret this result as
due to photon spawning (see e.g. \citealp{schonherr07}), suggesting this as the
likely explanation of the elusiveness of the fundamental. Moreover, 
\citet{furst14} report for the first time in this source a correlation between the
harmonic line energy and the flux, as expected in the sub-critical accretion 
regime \citep{becker12}.

In this paper  we performed a detailed spectral analysis of Vela X-1 
based on the eleven-year monitoring performed by the Burst Alert Telescope 
(BAT, \citealp{bat}) onboard \emph{Swift} \citep{swift}.
The large BAT field of view (1.4 steradian half coded), 
together with the \emph{Swift} observatory pointing strategy (several pointings  
per day towards different directions of the sky), allows BAT to have the source 
within its field of view nearly every day. 

This Paper is organized as follows. Section 2 describes the data reduction and the calibration
procedures applied to the BAT data. In Sect. 3 we describe the spectral analysis and
in Sect. 4 we discuss our results.

        \section{Data Reduction\label{data}}

The BAT survey data collected between 2004 December and 2015 November  
were retrieved from the HEASARC public
archive\footnote{http://heasarc.gsfc.nasa.gov/docs/archive.html} and 
 processed with the  {\sc bat\_imager} code \citep{segreto}, a software built for 
the analysis of data from coded mask instruments that performs screening, 
mosaicking and source detection and produces scientific products of any revealed 
source. 

Figure~\ref{lc} shows the light curve of Vela X-1 in the 15--100 keV band, with a
bin time equal to the orbital period (8.964 d). The intensity of the source shows 
large fluctuations, up to one order of magnitude.

The spectral analysis is aimed to investigate the behaviour of the
cyclotron features with luminosity. 
We produced background subtracted spectra of the source
at several count rate levels. The appropriate good time intervals (GTIs)
for each count rate interval were selected based on the source 15--150 keV light 
curve with a bin 
time of  90 minutes, in the count rate range between 0.001 and 0.05 counts 
s$^{-1}$ pixel$^{-1}$, corresponding to an observed 15-150 keV luminosity range
between $0.15$ and  $7.5\times10^{36}$ erg s$^{-1}$. To define the boundaries of
each count rate bin, we started from the lower count rate limit, and we 
increased the upper count rate limit with a step of 0.001 counts s$^{-1}$ 
pixel$^{-1}$ until a satisfactory signal-to-noise ratio (SNR) of $\sim 800$ is
reached, allowing for enough statistics for an adequate modeling of the 
second harmonic).
Since a Crab spectrum with similar statistics shows the presence of systematic
deviations with respect to a power law, we have applied to each Vela X-1 
spectrum a systematic error to be added in quadrature to the statistical one. 
The systematic errors were derived 
fitting the Crab spectrum extracted from the entire survey monitoring with a
power law and evaluating for each energy channel the values
$\rm |R_i -R^{\star}_i|/R_i$, where $R^{\star}_i$ is the rate derived with the 
best fitting powerlaw model in the i$_{th}$  channel and R$_i$ is the Crab count 
rate  in the same channel. We have verified that this correction (amounting
to 2\% on average) is 
negligible with respect to the statistical errors for the channels above 40 keV.
We used the official BAT spectral redistribution
matrix\footnote{http://heasarc.gsfc.nasa.gov/docs/heasarc/caldb/data/swift/bat/index.html}.
We report errors on spectral parameters at 68\,\% confidence level.

\section{Spectral analysis}
\label{spectra}

\begin{figure}
\begin{center}
\centerline{\includegraphics[width=9cm]{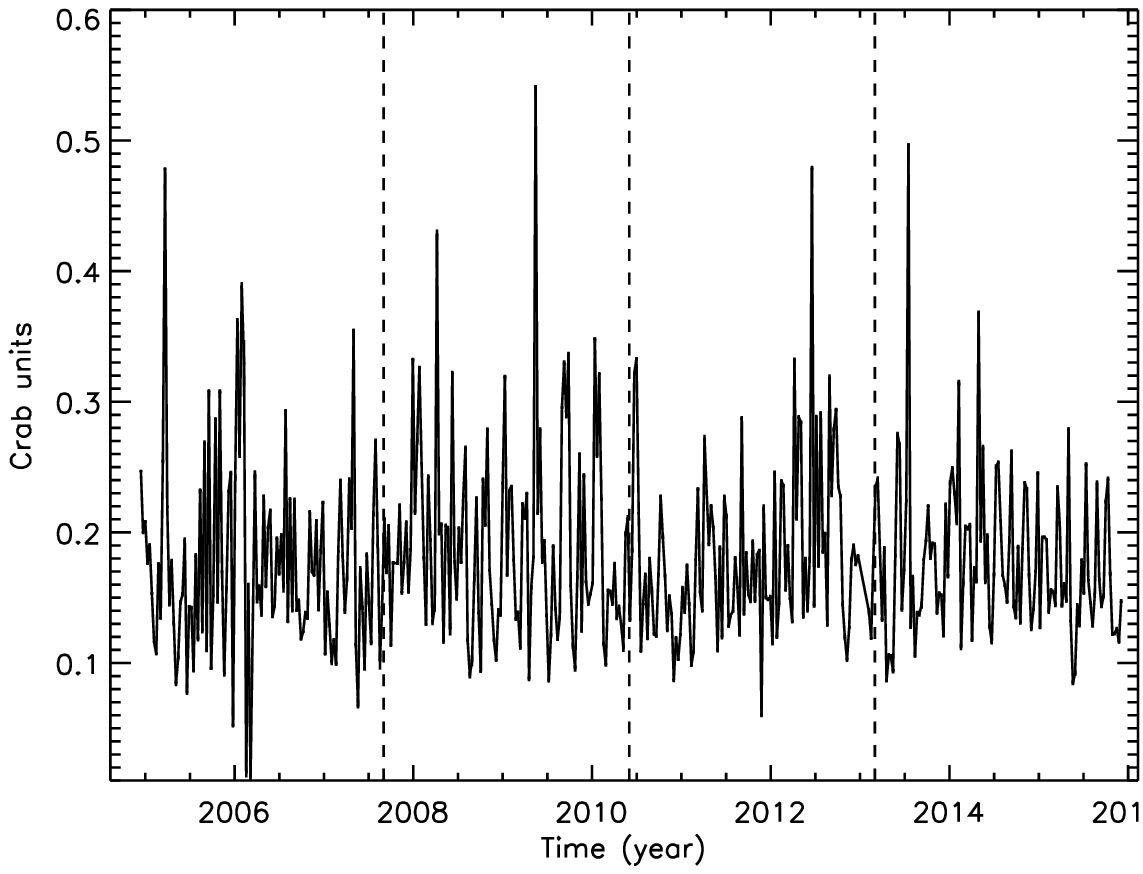}}

\caption[]{BAT 15-100 keV light curve of Vela X-1. The bin time is equal to 
the orbital period (8.964~d). The vertical dashed lines mark the four 
time-selected intervals used to probe the line ennergy vs luminosity relation.
}               
		\label{lc} 
        \end{center}
        \end{figure}

\begin{figure}
\begin{center}
\centerline{\includegraphics[width=9cm]{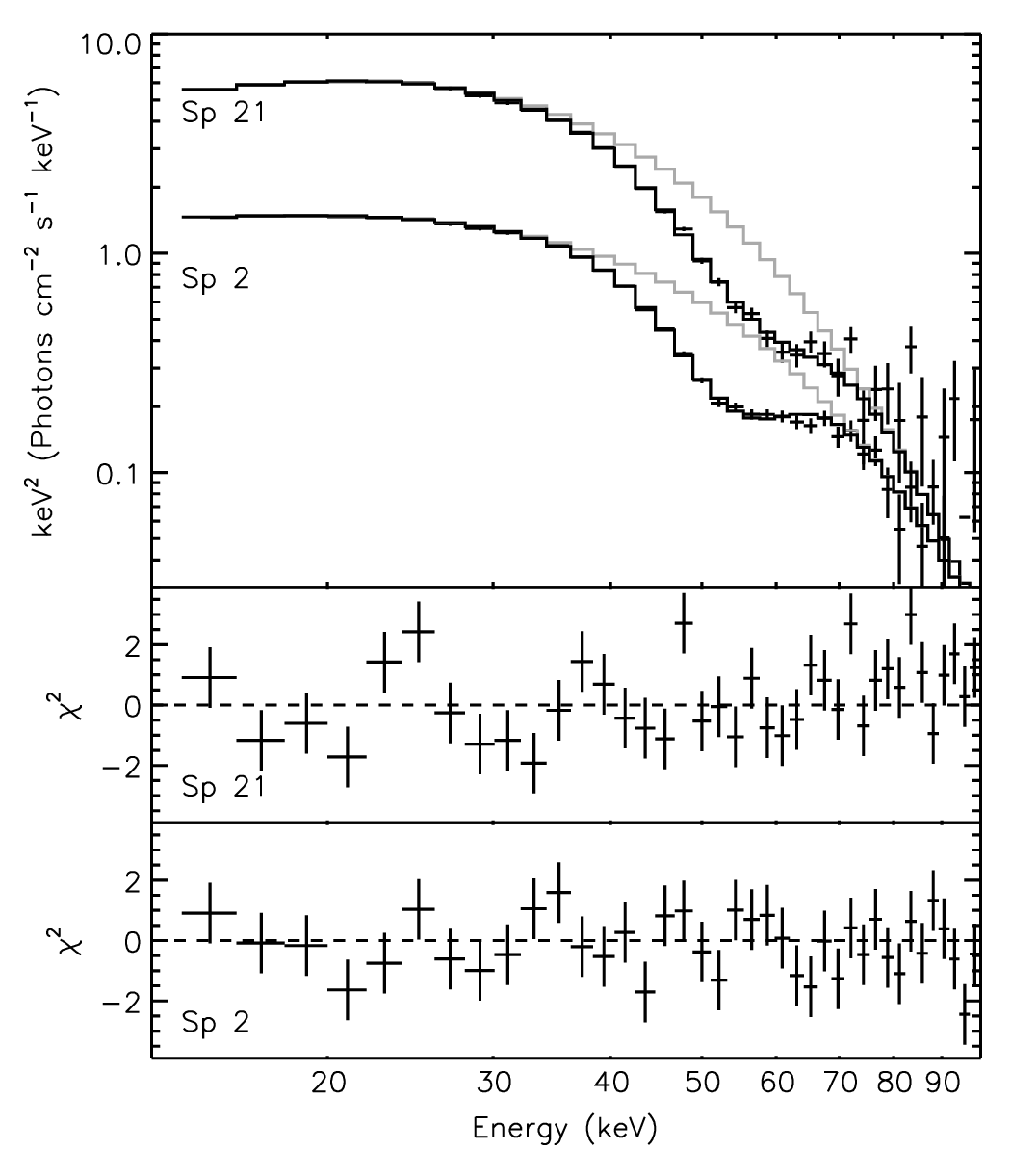}}

\caption[]{Data, best fit model, and residuals for two representative spectra
(spectra 2 and 21 in Table~\ref{sp_all}). In the top panel, the black line
represents the best fit model (continuum and CRSF), while  the grey line
represents the best fit continuum model.
}               
		\label{spettri} 
        \end{center}
        \end{figure}

\begin{figure}
\begin{center}
\centerline{\includegraphics[width=9cm]{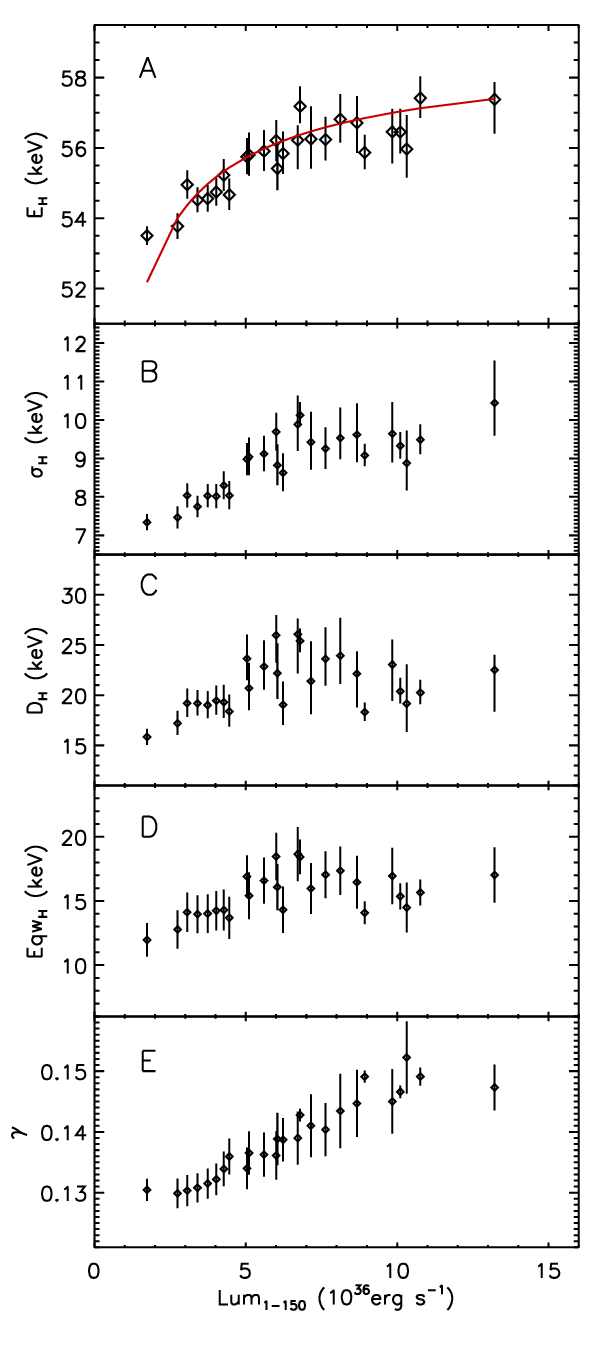}}

\caption[]{Best fit parameters vs source luminosity 
for the spectra extracted along the entire survey interval. From top to 
bottom: line energy,  width, strength, equivalent width,
Comptonization parameter. The solid line in the
top panel is the
theoreticical prediction for E$_{\star}$=29.56 keV (see Eq.~\ref{beck}).
}               
		\label{e0all} 
        \end{center}
        \end{figure}

To model the continuum emission of Vela X-1 we tested different spectral
shapes. A simple power law modified with an exponential cut-off ({\tt
cutoffpl}) is not adequate to describe the data. Both an
optically thick Comptonization ({\tt comptt}) and a Fermi-Dirac cutoff provide
an acceptable fit for the continuum emission, while the  CRSF is well described
with a Gaussian absorption profile ({\tt gabs} in Xspec).
We verified that the choice between the two 
continuum models does not affect the best fit line parameters. The results 
presented in this Paper refer to a continuum described with
an optically thick Comptonization.
The results of the spectral analysis are reported in Table 1.  
In all the spectra we detect with high significance the second harmonic, while
the fundamental cannot be detected in any of the spectra. Figure~\ref{spettri}
shows two representative spectra at different levels of intensity, and their
residuals with respect to the relevant best fit model.
Figure~\ref{e0all} shows the best fit line parameters (energy, width, strength
and equivalent width in panels A, B, C, D, respectively) as a function of the 
intrinsic source luminosity. 
The equivalent width was evaluated according to the definition:
\begin{equation}
\rm EW=\int{1-\frac{F(E)}{F_C(E)}dE}
\end{equation}
where $\rm F_C$ represents the intensity of the continuum and $\rm F$ is the 
intensity of the best fit model, evaluated in the  15--100 keV energy range. 
The line energy shows a clear direct  correlation with the luminosity, ranging
from $\sim 53$ keV at low luminosity to $\sim 58$ keV at high luminosity. The
other line parameters also  show a positive correlation with the luminosity.
Panel E of figure~\ref{e0all} shows the Comptonization
parameter, defined as $\gamma=\tau kT/m_ec^2$. We observe that $\gamma$ 
is positively correlated with the
luminosity, indicating a hardening of the continuum at higher luminosity, 
as already reported  by \citet{odaka13} and \citet{furst14} for Vela X-1 and by
\citet{klochkov11} for other accreting pulsars in subcritical accretion regime.

\begin{figure}
\begin{center}
\centerline{\includegraphics[width=9cm]{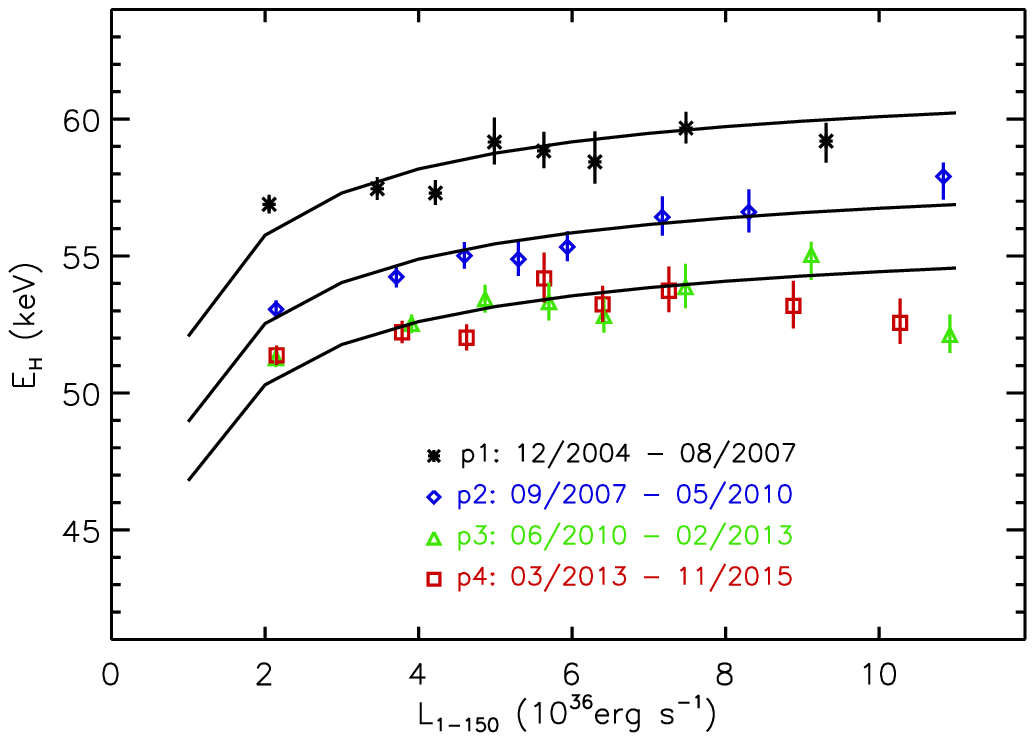}}

\caption[]{Energy of the second harmonic vs source luminosity 
for the four 33-month time intervals. The continuous lines in the top panel are 
theoretical predictions (see Eq.~\ref{beck}) for E$_{\star}$=28.25, 29.42, 31.12 keV 
(bottom to top, respectively). The E$_{\star}$ values were obtained fitting each set
of data with Equation~\ref{beck}. Data sets p3 and p4 were fitted together. 
}               
		\label{e0time} 
        \end{center}
        \end{figure}

We have also investigated the long-term behaviour of this trend in time selected
intervals. To this aim
we have split the survey monitoring into four time intervals of the same length
(33 months, see dashed lines in Fig.~\ref{lc}), selecting the spectra in eight 
flux intervals for each of them. The
results of the spectral analysis are plotted in Figure~\ref{e0time}. We find
that the line energy shows a systematic decrease between the first and the second
time interval and between the second and the third one, while there is no
significant change between the third and the fourth interval.

\section{Discussion and conclusions\label{discuss}}
We report on the analysis of the Swift-BAT spectral data of Vela X-1, 
focussing on the on the relation between the CRSF energy and luminosity in 
the overall timespan covered by BAT and in time-selected intervals. 

The energy of the second
harmonic is directly correlated with the luminosity  (with a correlation
coefficient of 0.89, and a corresponding probability of no correlation of
$\sim1.0\times 10^{-9}$) , confirming the results
reported by \citet{furst14} from the analysis of two NuSTAR observations. This 
correlation, observed for the first time in Her X-1 \citep{staubert07} is 
expected in the sub-critical accretion regime
($L_{coul}<L<L_{crit}$, see \citealp{becker12}),
where the deceleration of the material to rest at the stellar surface is
accomplished by a shock dominated by Coulomb interactions. The ram pressure of
the infalling plasma increases with the accretion rate driving the shock region
down to regions of higher magnetic field intensity. Using equations (51) and
(58) in  \citet{becker12}, we can derive the energy/luminosity
correlation for the fundamental CRSF:
\begin{multline}
E_{\rm{F}} =   \Biggl[1+ 0.6  \left(\frac{R{_\star}}{10\,\rm{km}}\right)^{-\frac{13}{14}}  \left(\frac{\Lambda}{0.1}\right)^{-1} \left(\frac{\tau_{\star}}{20}\right)
\left(\frac{M_{\star}}{1.4\,\rm M_{\odot}}\right)^{\frac{19}{14}} \\
\times  \left(\frac{E_{\star}}{1\,\rm{keV}}\right)^{-\frac{4}{7}}\left(\frac{L_x}{10^{37}\,\rm{erg\,s}^{-1}}\right)^{-\frac{5}{7}}\Biggr]^{-3}\times E_{\star}\quad\rm{.}
\label{beck}
\end{multline}
where $\tau_\star$ is the Thomson optical depth ( $\sim20$ for 
typical HMXB parameters, \citealp{becker12}), $R_{\star}$ is the radius of the
NS, $M_{\star}$ is the mass of the NS, $E_{\star}$ is the energy of the
fundamental cyclotron line at the NS surface and $\Lambda$ is a constant related
to the interaction between the magnetic field and the surrounding medium
\citep{lamb73}. 
We adopt the values $\Lambda=1$, suitable for spherical accretion, $R_{\star}=10$ km
and $M_{\star}=1.8 M_{\odot}$ \citep{rawls11} as in 
\citet{furst14}. The energy of the CRSF depends on the accretion rate in 
good agreement with theoretical predictions: a fit of the data using as best fit
model Equation~\ref{beck} yields $E_{\star}=29.56\pm0.05$, 
with $\chi^2=1.8$, assuming a harmonic ratio of 2 (solid line in 
the top panel of Fig~\ref{e0all}). 

\citet{mushtukov15} have proposed an alternative theory to explain the positive 
correlation between CRSF energy and luminosity in subcritical sources: the bulk
of the radiation, emitted by the hotspot on the polar cap, travels upwards
through the accretion channel interacting with the infalling matter and
producing the CRSF at the resonant energies. The line appears redshifted by 
Doppler effect  to the observer due to the velocity profile of the plasma 
falling towards the NS surface. The higher the luminosity, the lower is the
velocity of plasma, slowed by the radiation pressure, and the smaller is the
Doppler shift of the line. At the limit of the critical luminosity that marks the
balance between radiation pressure and ram pressure of the matter, no Doppler
shift is observed and the line energy corresponds to the value defined by the
surface magnetic field. In this framework, the width of the line is also
expected to correlate positively with the luminosity, since 
an increase of the radiation pressure produces a layer of infalling matter at lower
velocity, extending blueward the line profile. These trends are indeed observed 
in our results (Fig~\ref{e0all})


We have investigated if this correlation evolves on a yearly time scale, 
splitting the survey data
into four 33-month intervals. We observe a significant shift in energy of this 
correlation 
along the time, except for the last two intervals, where the measurements
overlap. The
correlation coefficient for each data set are 0.83 (p-value=0.010), 0.99
(p-value=$4.3\times10^{-6}$), 0.47 (p-value=0.066) for p1, p2 and p3+p4,
respectively.
The theoretical correlation matches the observed data for different values of
the energy of the fundamental at the stellar surface ($28.25\pm0.07$ keV,
$29.42\pm0.09$ keV and $31.12\pm0.10$ keV, with $\chi^2$ of 2.3, 1.2, and 2.4,
respectively, solid lines in the top panel of Figure~\ref{e0time}). 
This translates into 
a decrease of the fundamental line of $\sim0.36$ keV/year along the first three time
intervals, corresponding to a decay of the surface magnetic field intensity of 
$\sim3\times 10^{10}$ G/year.

We have checked if this shift could be due to a systematic drift of the
instrumental energy/channel gain. Such a drift would produce a significant
variation in the slope of the Crab spectrum along the time. Therefore we
accumulated the Crab spectra in the four time intervals and 
fitted them with a power-law. The variation of the Crab power-law photon
index between the last
and the first time interval is $+0.017\pm 0.006$. If this was due to a variation in
the channel/energy gain, this would correspond to an uncertainty in the energy
determination of only $\sim 0.2$ keV, negligible with respect to the statistical
uncertainty on the line position.

A long-term decay  in the magnetic field, unrelated to the source luminosity, 
was recently assessed in the 
persistent accreting pulsar Her X-1 \citep{staubert14,staubert16}, where the cyclotron 
line energy decreased on average by 0.26 keV/year between 1996 and 2012. 
This long term decay is most likely a local effect confined to the magnetic 
polar cap: it could be related to the accreted matter that accumulates into a
magnetically supported mound and causes either 
a distortion of the magnetic field lines \citep{brown98} or a geometrical 
displacement of the emission region. The observed break in the decay after the
third time interval could be explained if this mound has reached a maximum size for
a stable structure \citep{mukherjee12,litwin01} and the plasma settling on it 
is balanced  by the plasma leaking out from its base. A similar interpretation
was adopted to explain the drop of the magnetic field in V0332+53 during the
2015 outburst, where a difference of $\sim1.5$ keV in the CRSF fundamental line
energy was observed between the start and the end of the outburst
\citep{cusumano16}.

With an average luminosity of $\sim 3.0\times 10^{36}$ erg/s Vela X-1 accretes 
at $\sim 0.36 \times 10^{-9} M_{\odot}$ per year, while the average 
luminosity of $\sim 10^{37}$ erg/s in Her X-1 corresponds to 
$\sim 10^{-9} M_{\odot}$ per year \citep{klochkov15}.
If the accreted matter is the cause of the decay of the magnetic field 
along the accretion column, this mechanism is a factor of $\sim4$ more 
efficient in Vela X-1 than in Her X-1. This difference could
be related to a narrower accretion column in Vela X-1, that, following Eq. 23 in
\citet{becker12}, is $\sim 0.4$ km with respect to  $\sim 1.3$ km in Her X-1 and
allows for a faster growth of the height of the mound. Indeed, the accreted mass
per unit surface is $\sim 7.2\times10^{-10} M_{\sun}$ km$^{-2}$ year$^{-1}$ in
Vela X-1, a factor of $\sim4$ higher than $\sim 1.9\times10^{-10} M_{\sun}$ 
km$^{-2}$ year$^{-1}$ in Her X-1.

As shown in Figure 4 in \citet{staubert14}, the CRSF energy in Her X-1 was
constant before 1991. Between 1991 and 1995 a drop of $\sim 8$ kev was observed,
followed by a more linear decay after
1995. One explanation might be a cyclic behaviour of the CRSF energy, 
and the jump could be induced by a destructive event that destroys the mound 
and resets the line emission region to the pre-mound configuration. Vela X-1 
could have entered a stable configuration in the last time interval. An abrupt
recovery of the line energy to a higher value in the next years could be 
therefore expected if the same restoring mechanism observed in Her X-1 works 
also in Vela X-1.


\begin{table}
\renewcommand{\arraystretch}{1.5}
\scriptsize
\begin{tabular}{r p{0.8cm} p{0.8cm} p{0.8cm} p{0.8cm} p{0.8cm} p{0.9cm} r}
\hline
Sp & $E_{\rm H}$    & $\sigma_{\rm H}$ & $D_{\rm H}$     &   $kT$         &   $\tau$             & $L_{\rm 36}$     & $\chi^2$\\
   & keV            &  keV             &                 &   keV          &                      & erg s$^{-1}$     & \\
\hline 
   1 & $53.4_{-0.3}^{+0.3}$ & $ 7.3_{-0.2}^{+0.2}$ & $15.3_{-0.9}^{+0.9}$ & $ 8.5_{-0.1}^{+0.1}$ & $ 7.8_{-0.1}^{+0.1}$ & 1.73 & 1.27 \\
   2 & $53.8_{-0.4}^{+0.4}$ & $ 7.5_{-0.3}^{+0.3}$ & $17.2_{-1.2}^{+1.2}$ & $ 8.5_{-0.1}^{+0.1}$ & $ 7.8_{-0.1}^{+0.1}$ & 2.74 & 1.10 \\
   3 & $55.0_{-0.4}^{+0.4}$ & $ 8.0_{-0.3}^{+0.3}$ & $19.2_{-1.4}^{+1.5}$ & $ 8.3_{-0.1}^{+0.1}$ & $ 8.0_{-0.1}^{+0.1}$ & 3.06 & 1.26 \\
   4 & $54.5_{-0.3}^{+0.4}$ & $ 7.7_{-0.3}^{+0.3}$ & $19.2_{-1.2}^{+1.3}$ & $ 8.3_{-0.1}^{+0.1}$ & $ 8.0_{-0.1}^{+0.1}$ & 3.40 & 1.44 \\
   5 & $54.6_{-0.4}^{+0.4}$ & $ 8.0_{-0.3}^{+0.3}$ & $19.0_{-1.3}^{+1.4}$ & $ 8.2_{-0.1}^{+0.1}$ & $ 8.2_{-0.1}^{+0.1}$ & 3.74 & 1.12 \\
   6 & $54.7_{-0.4}^{+0.4}$ & $ 8.0_{-0.3}^{+0.3}$ & $19.4_{-1.4}^{+1.5}$ & $ 8.1_{-0.1}^{+0.1}$ & $ 8.4_{-0.1}^{+0.1}$ & 4.02 & 1.15 \\
   7 & $55.2_{-0.4}^{+0.5}$ & $ 8.3_{-0.4}^{+0.4}$ & $19.3_{-1.6}^{+1.7}$ & $ 7.9_{-0.1}^{+0.1}$ & $ 8.6_{-0.1}^{+0.1}$ & 4.27 & 0.78 \\
   8 & $54.7_{-0.4}^{+0.5}$ & $ 8.0_{-0.4}^{+0.4}$ & $18.4_{-1.5}^{+1.7}$ & $ 7.7_{-0.1}^{+0.1}$ & $ 9.0_{-0.2}^{+0.2}$ & 4.45 & 0.99 \\
   9 & $55.8_{-0.5}^{+0.5}$ & $ 9.0_{-0.4}^{+0.4}$ & $24_{-2}^{+2}$       & $ 8.0_{-0.1}^{+0.1}$ & $ 8.5_{-0.2}^{+0.2}$ & 5.03 & 0.70 \\
  10 & $55.8_{-0.6}^{+0.6}$ & $ 9.0_{-0.5}^{+0.5}$ & $21_{-2}^{+2}$       & $ 7.7_{-0.1}^{+0.1}$ & $ 9.0_{-0.2}^{+0.2}$ & 5.10 & 1.24 \\
  11 & $55.9_{-0.6}^{+0.6}$ & $ 9.1_{-0.4}^{+0.5}$ & $23_{-2}^{+3}$       & $ 7.8_{-0.1}^{+0.1}$ & $ 9.0_{-0.2}^{+0.2}$ & 5.60 & 1.04 \\
  12 & $56.2_{-0.6}^{+0.6}$ & $ 9.7_{-0.5}^{+0.5}$ & $26_{-3}^{+2}$       & $ 7.9_{-0.2}^{+0.2}$ & $ 8.8_{-0.2}^{+0.2}$ & 6.00 & 1.15 \\
  13 & $55.4_{-0.6}^{+0.7}$ & $ 8.8_{-0.5}^{+0.5}$ & $22_{-2}^{+3}$       & $ 7.6_{-0.1}^{+0.2}$ & $ 9.4_{-0.2}^{+0.2}$ & 6.04 & 1.33 \\
  14 & $55.8_{-0.6}^{+0.6}$ & $ 8.6_{-0.5}^{+0.5}$ & $19_{-2}^{+2}$       & $ 7.5_{-0.1}^{+0.1}$ & $ 9.5_{-0.2}^{+0.2}$ & 6.23 & 1.38 \\
  15 & $56.2_{-0.8}^{+0.4}$ & $ 9.9_{-0.7}^{+0.7}$ & $26_{-4}^{+2}$       & $ 7.7_{-0.2}^{+0.1}$ & $ 9.3_{-0.2}^{+0.3}$ & 6.71 & 1.36 \\
  16 & $57.2_{-0.5}^{+0.6}$ & $10.1_{-0.3}^{+0.3}$ & $25.4_{-1.1}^{+1.2}$ & $ 7.5_{-0.1}^{+0.1}$ & $ 9.8_{-0.1}^{+0.1}$ & 6.79 & 1.34 \\
  17 & $56.3_{-0.8}^{+0.9}$ & $ 9.4_{-0.7}^{+0.8}$ & $21_{-3}^{+4}$       & $ 7.5_{-0.2}^{+0.2}$ & $ 9.7_{-0.3}^{+0.3}$ & 7.15 & 1.04 \\
  18 & $56.2_{-0.6}^{+0.6}$ & $ 9.3_{-0.5}^{+0.6}$ & $24_{-3}^{+3}$       & $ 7.3_{-0.1}^{+0.2}$ & $ 9.8_{-0.2}^{+0.2}$ & 7.63 & 1.77 \\
  19 & $56.8_{-0.7}^{+0.7}$ & $ 9.5_{-0.5}^{+0.8}$ & $24_{-3}^{+4}$       & $ 7.3_{-0.1}^{+0.2}$ & $10.1_{-0.4}^{+0.2}$ & 8.12 & 0.80 \\
  20 & $56.7_{-0.9}^{+0.8}$ & $ 9.6_{-0.7}^{+0.8}$ & $22_{-3}^{+2}$       & $ 7.1_{-0.1}^{+0.1}$ & $10.5_{-0.3}^{+0.3}$ & 8.67 & 1.09 \\
  21 & $55.9_{-0.5}^{+0.5}$ & $ 9.1_{-0.3}^{+0.3}$ & $18.3_{-0.9}^{+0.9}$ & $ 7.0_{-0.1}^{+0.1}$ & $10.9_{-0.1}^{+0.1}$ & 8.92 & 1.92 \\
  22 & $56.5_{-0.9}^{+0.7}$ & $ 9.6_{-0.7}^{+0.8}$ & $23.0_{-4}^{+2}$     & $ 7.0_{-0.2}^{+0.1}$ & $10.6_{-0.3}^{+0.3}$ & 9.84 & 1.48 \\
  23 & $56.5_{-0.6}^{+0.7}$ & $ 9.3_{-0.3}^{+0.4}$ & $20.4_{-1.2}^{+1.3}$ & $ 6.9_{-0.1}^{+0.1}$ & $10.9_{-0.1}^{+0.1}$ &10.10 & 1.26 \\
  24 & $56.0_{-0.8}^{+1.0}$ & $ 8.9_{-0.7}^{+0.8}$ & $19_{-3}^{+4}$       & $ 6.8_{-0.1}^{+0.2}$ & $11.5_{-0.4}^{+0.4}$ &10.31 & 1.37 \\
  25 & $57.4_{-0.6}^{+0.6}$ & $ 9.5_{-0.4}^{+0.4}$ & $20.2_{-1.1}^{+1.2}$ & $ 6.8_{-0.1}^{+0.1}$ & $11.2_{-0.1}^{+0.1}$ &10.76 & 1.47 \\
  26 & $57.4_{-1.0}^{+0.5}$ & $10.4_{-0.8}^{+1.1}$ & $22_{-4}^{+1}$ & $ 6.8_{-0.2}^{+0.1}$ & $11.1_{-0.1}^{+0.3}$ &13.21 & 1.01 \\
\hline
\end{tabular}
\caption{Spectral fit results from the count-rate selected spectra extracted 
along the entire  survey monitoring. E$_H$, $\sigma_H$ and D$_H$ are the 
energy, width and depth of the CRSF second harmonic, respectively. kT and $\tau$
are the plasma temperature and optical depth, respectively. The input photon
temperature attains to the soft X-ray energy range and it is not well constrained
by the BAT data. Therefore, it has been kept fixed to 0.82 keV for all the 
spectra, which is the average value obtained in a pre-analysis of the spectra.
$L_{36}$ is the luminosity of the continuum evaluated in the 1-150 keV energy 
range in units of $10^{36}$ erg s$^{-1}$. The quoted reduced $\chi^2$ is 
evaluated for 32 degrees of freedom.
\label{sp_all}}
\end{table}

\section*{Acknowledgments}

This work was supported by contract ASI I/004/11/0.
We thank the anonymous referee for comments that helped improve the paper. 


\begin{table}
\renewcommand{\arraystretch}{1.5}
\scriptsize
\begin{tabular}{r p{0.8cm} p{0.8cm} p{0.8cm} p{0.8cm} p{0.8cm} p{0.8cm}  r}
\hline
Sp & $E_{\rm H}$    & $\sigma_{\rm H}$ & $D_{\rm H}$     &   $kT$         &   $\tau$             & $L_{\rm 36}$     & $\chi^2$\\
   & keV            &  keV       &                 &   keV          &                      & erg s$^{-1}$     & \\
\hline 
   p1-1 & $56.9_{-0.3}^{+0.3}$ & $ 7.9_{-0.3}^{+0.3}$ & $20.1_{-1.1}^{+1.2}$ & $ 9.0_{-0.1}^{+0.1}$ & $ 7.6_{-0.1}^{+0.1}$ & 2.05 & 1.66 \\
   p1-2 & $57.5_{-0.4}^{+0.4}$ & $ 8.3_{-0.3}^{+0.3}$ & $20.9_{-1.4}^{+1.6}$ & $ 8.5_{-0.1}^{+0.1}$ & $ 8.2_{-0.1}^{+0.1}$ & 3.46 & 1.42 \\
   p1-3 & $57.3_{-0.4}^{+0.5}$ & $ 8.1_{-0.4}^{+0.4}$ & $19.9_{-1.5}^{+1.7}$ & $ 8.1_{-0.1}^{+0.1}$ & $ 8.9_{-0.1}^{+0.1}$ & 4.22 & 1.30 \\
   p1-4 & $59.2_{-0.8}^{+0.9}$ & $ 9.4_{-0.6}^{+0.7}$ & $24_{-3}^{+3}$       & $ 8.0_{-0.1}^{+0.2}$ & $ 9.2_{-0.2}^{+0.2}$ & 4.99 & 1.21 \\
   p1-5 & $58.8_{-0.6}^{+0.7}$ & $ 9.0_{-0.5}^{+0.6}$ & $23_{-2}^{+3}$       & $ 7.8_{-0.1}^{+0.1}$ & $ 9.4_{-0.2}^{+0.2}$ & 5.63 & 1.19 \\
   p1-6 & $58.4_{-0.8}^{+1.1}$ & $ 9.6_{-0.7}^{+0.8}$ & $22_{-3}^{+3}$       & $ 7.6_{-0.1}^{+0.1}$ & $10.2_{-0.2}^{+0.3}$ & 6.29 & 1.80 \\
   p1-7 & $61.2_{-0.8}^{+0.5}$ & $11.1_{-1.4}^{+0.5}$ & $32.8_{-8.2}^{+0.9}$ & $ 7.6_{-0.3}^{+0.1}$ & $10.3_{-0.1}^{+0.1}$ & 7.73 & 1.66 \\
   p1-8 & $59.2_{-0.8}^{+0.7}$ & $ 9.7_{-0.7}^{+0.9}$ & $24_{-3}^{+2}$       & $ 7.2_{-0.2}^{+0.1}$ & $11.1_{-0.3}^{+0.4}$ & 9.31 & 1.48 \\
\hline
   p2-1 & $53.1_{-0.3}^{+0.3}$ & $ 6.7_{-0.3}^{+0.3}$ & $15.7_{-1.0}^{+1.0}$ & $ 8.3_{-0.1}^{+0.1}$ & $ 8.0_{-0.1}^{+0.1}$ & 2.14 & 1.19 \\
   p2-2 & $54.2_{-0.4}^{+0.4}$ & $ 7.5_{-0.3}^{+0.3}$ & $18.6_{-1.3}^{+1.4}$ & $ 8.1_{-0.1}^{+0.1}$ & $ 8.2_{-0.1}^{+0.1}$ & 3.71 & 1.71 \\
   p2-3 & $55.0_{-0.5}^{+0.5}$ & $ 8.4_{-0.4}^{+0.4}$ & $21.8_{-1.9}^{+2.1}$ & $ 8.0_{-0.1}^{+0.1}$ & $ 8.6_{-0.2}^{+0.2}$ & 4.60 & 0.77 \\
   p2-4 & $54.9_{-0.6}^{+0.7}$ & $ 8.3_{-0.5}^{+0.5}$ & $19_{-2}^{+3}$       & $ 7.6_{-0.1}^{+0.2}$ & $ 9.2_{-0.2}^{+0.2}$ & 5.30 & 0.92 \\
   p2-5 & $55.3_{-0.5}^{+0.6}$ & $ 8.4_{-0.4}^{+0.5}$ & $21_{-2}^{+2}$       & $ 7.5_{-0.1}^{+0.1}$ & $ 9.4_{-0.2}^{+0.2}$ & 5.93 & 0.96 \\
   p2-6 & $56.4_{-0.7}^{+0.8}$ & $ 9.4_{-0.6}^{+0.6}$ & $25_{-3}^{+4}$       & $ 7.6_{-0.2}^{+0.2}$ & $ 9.3_{-0.3}^{+0.3}$ & 7.17 & 1.00 \\
   p2-7 & $56.6_{-0.7}^{+0.8}$ & $ 9.0_{-0.6}^{+0.7}$ & $22_{-3}^{+3}$       & $ 7.2_{-0.1}^{+0.2}$ & $10.2_{-0.3}^{+0.3}$ & 8.30 & 0.72 \\
   p2-8 & $57.9_{-0.8}^{+0.5}$ & $10.3_{-0.7}^{+0.9}$ & $28_{-4}^{+2}$       & $ 7.3_{-0.2}^{+0.1}$ & $10.0_{-0.2}^{+0.3}$ &10.83 & 1.12 \\
\hline
   p3-1 & $51.3_{-0.3}^{+0.3}$ & $ 6.7_{-0.3}^{+0.3}$ & $14.3_{-0.9}^{+1.0}$ & $ 8.2_{-0.1}^{+0.1}$ & $ 3.5_{-0.1}^{+0.1}$ & 2.14 & 1.53 \\
   p3-2 & $52.5_{-0.3}^{+0.3}$ & $ 7.2_{-0.3}^{+0.3}$ & $18.3_{-1.2}^{+1.3}$ & $ 8.0_{-0.1}^{+0.1}$ & $ 8.0_{-0.1}^{+0.1}$ & 3.91 & 1.11 \\
   p3-3 & $53.4_{-0.5}^{+0.5}$ & $ 8.1_{-0.4}^{+0.4}$ & $20.0_{-1.9}^{+2.0}$ & $ 7.7_{-0.1}^{+0.1}$ & $ 8.6_{-0.2}^{+0.2}$ & 4.87 & 0.67 \\
   p3-4 & $53.3_{-0.7}^{+0.7}$ & $ 8.4_{-0.5}^{+0.5}$ & $22_{-3}^{+3}$       & $ 7.6_{-0.2}^{+0.2}$ & $ 8.7_{-0.2}^{+0.2}$ & 5.70 & 1.27 \\
   p3-5 & $52.8_{-0.6}^{+0.7}$ & $ 8.0_{-0.5}^{+0.6}$ & $19_{-2}^{+3}$       & $ 7.3_{-0.2}^{+0.2}$ & $ 9.3_{-0.3}^{+0.3}$ & 6.41 & 1.40 \\
   p3-6 & $53.9_{-0.8}^{+0.9}$ & $ 9.2_{-0.6}^{+0.7}$ & $24_{-3}^{+4}$       & $ 7.3_{-0.2}^{+0.2}$ & $ 9.5_{-0.3}^{+0.3}$ & 7.47 & 0.77 \\
   p3-7 & $55.0_{-0.9}^{+0.5}$ & $10.0_{-0.8}^{+0.8}$ & $29_{-5}^{+1}$       & $ 7.3_{-0.2}^{+0.1}$ & $ 9.5_{-0.2}^{+0.4}$ & 9.11 & 1.22 \\
   p3-8 & $52.1_{-0.7}^{+0.7}$ & $ 7.4_{-0.6}^{+0.7}$ & $15_{-2}^{+3}$       & $ 6.4_{-0.1}^{+0.1}$ & $11.4_{-0.4}^{+0.4}$ &10.92 & 1.23 \\
\hline
   p4-1 & $51.4_{-0.3}^{+0.4}$ & $ 6.9_{-0.3}^{+0.3}$ & $15.0_{-1.0}^{+1.1}$ & $ 8.2_{-0.1}^{+0.1}$ & $ 8.0_{-0.1}^{+0.1}$ & 2.15 & 1.53 \\
   p4-2 & $52.2_{-0.4}^{+0.4}$ & $ 7.6_{-0.3}^{+0.3}$ & $18.2_{-1.4}^{+1.5}$ & $ 7.9_{-0.1}^{+0.1}$ & $ 8.4_{-0.1}^{+0.1}$ & 3.78 & 1.08 \\
   p4-3 & $52.0_{-0.5}^{+0.5}$ & $ 7.1_{-0.4}^{+0.4}$ & $15.1_{-1.4}^{+1.6}$ & $ 7.3_{-0.1}^{+0.1}$ & $ 9.2_{-0.2}^{+0.2}$ & 4.62 & 0.57 \\
   p4-4 & $54.2_{-0.9}^{+0.9}$ & $ 9.3_{-0.7}^{+0.7}$ & $23_{-4}^{+4}$       & $ 7.7_{-0.2}^{+0.2}$ & $ 8.8_{-0.3}^{+0.3}$ & 5.63 & 1.18 \\
   p4-5 & $53.2_{-0.6}^{+0.7}$ & $ 8.7_{-0.5}^{+0.5}$ & $22_{-3}^{+3}$       & $ 7.5_{-0.2}^{+0.2}$ & $ 9.1_{-0.2}^{+0.2}$ & 6.39 & 1.22 \\
   p4-6 & $53.7_{-0.8}^{+0.9}$ & $ 8.6_{-0.7}^{+0.7}$ & $20_{-3}^{+4}$       & $ 7.1_{-0.2}^{+0.2}$ & $ 9.8_{-0.3}^{+0.3}$ & 7.26 & 0.94 \\
   p4-7 & $53.2_{-0.8}^{+0.9}$ & $ 8.7_{-0.7}^{+0.8}$ & $19_{-3}^{+4}$       & $ 6.9_{-0.2}^{+0.2}$ & $10.2_{-0.4}^{+0.4}$ & 8.88 & 0.97 \\
   p4-8 & $52.6_{-0.8}^{+0.9}$ & $ 7.6_{-0.7}^{+0.8}$ & $14_{-2}^{+3}$       & $ 6.5_{-0.1}^{+0.2}$ & $11.4_{-0.4}^{+0.4}$ &10.27 & 1.38 \\
\hline
\end{tabular}
\caption{Spectral fit results from the count-rate selected spectra extracted 
along the four 33-month time intervals. The columns are described in
Table~\ref{sp_all}.  \label{sp_p4}}
\end{table}

\bsp

\label{lastpage}

\end{document}